\documentclass[conference]{IEEEtran}
\IEEEoverridecommandlockouts

\usepackage{cite}
\usepackage{amsmath,amssymb,amsfonts}
\usepackage{algorithmic}
\usepackage{graphicx}
\usepackage{textcomp}
\usepackage{xcolor}
\usepackage{booktabs}
\usepackage{microtype}
\usepackage{url}
\usepackage{multirow}

\def\BibTeX{{\rm B\kern-.05em{\sc i\kern-.025em b}\kern-.08em
    T\kern-.1667em\lower.7ex\hbox{E}\kern-.125emX}}

\makeatother

\newcommand{\system}{MLN-WavLM}
\newcommand{\syspca}{\system{}-PCA}
\newcommand{\sysfull}{\system{}-Full}
\newcommand{\supref}{WavLM-HM-Conformer}

\begin{document}
\IEEEtriggeratref{12}

\title{When Normality Shifts: Multi-Level WavLM Modeling for Open-World Audio Deepfake Detection}

\author{
\IEEEauthorblockN{Daniel Niekurzak\IEEEauthorrefmark{1}, Daniel Kozuch\IEEEauthorrefmark{1}, Weronika Folwarska\IEEEauthorrefmark{2}, Khan Muhammad\IEEEauthorrefmark{3}, Rafał Scherer\IEEEauthorrefmark{1}\IEEEauthorrefmark{4}}
\IEEEauthorblockA{\IEEEauthorrefmark{1}Department of Artificial Intelligence,
Czestochowa University of Technology,
Poland\\
\texttt{\{daniel.niekurzak,daniel.kozuch,rafal.scherer\}@pcz.pl}}
\IEEEauthorblockA{\IEEEauthorrefmark{2}Department of Computer Science, 
Czestochowa University of Technology,
Poland\\
\texttt{weronika.folwarska@pcz.pl}}
\IEEEauthorblockA{\IEEEauthorrefmark{3}School of Convergence, Sungkyunkwan University, Seoul 03063, Republic of Korea\\
\texttt{khan.muhammad@ieee.org}}
\IEEEauthorblockA{\IEEEauthorrefmark{4}Faculty of Computer Science, AGH University of Krakow, Poland\\
\texttt{rafal.scherer@agh.edu.pl}}
}

\maketitle

\begin{abstract}
Audio deepfake detectors trained with known synthesis attacks can learn generator-specific shortcuts that fail as synthesis systems and acoustic conditions evolve. 
We investigate an attack-free alternative: a normality model trained exclusively on bona fide speech, which scores synthetic audio by its deviation from authentic representations. We propose \system{}, a method that applies shrinkage-Mahalanobis modeling to pooled hidden states from a frozen large-scale, pre-trained speech processing WavLM Base+ architecture.
Evaluated on ASVspoof 2019 LA, our best attack-free variant achieves an 8.45\% Equal Error Rate (EER), closely tracking a fully supervised WavLM-HM-Conformer baseline (7.86\%). However, when subjected to the ASVspoof 2021 DF domain shift, performance drops to 22.12\% EER, exposing vulnerabilities in the bona-fide-calibrated operating point. Layer-wise analysis reveals that early WavLM representations dominate the ensemble. To test real-world viability, we simulated deployment streams using unlabeled threshold adaptation. While robust median/MAD recalibration easily handles in-domain contamination up to 10\%, recalibrating under the 2021 DF domain shift triggers a total collapse of low-FPR recall. We conclude that while frozen source-domain normality can effectively rank deepfakes, it does not yield a stable open-world decision rule, and downstream threshold adaptation cannot rescue upstream ranking degradation.
\end{abstract}


\section{Introduction}
Modern text-to-speech and voice-conversion systems can synthesize convincing speech from short references, while the generators and channels encountered after deployment are rarely covered completely by labelled training attacks. This is an open-world detection problem: both the anomaly mechanisms and the acoustic domain can evolve. Supervised countermeasures remain strong when representative spoof data are available, but they can exploit artifacts of known generators and codecs.

Anomaly detection offers a different operational assumption. Instead of learning a boundary between authentic speech and a finite attack set, a model estimates the distribution of bona fide speech and assigns a high score to observations that depart from it. That setup is useful when new synthesis methods appear faster than labelled countermeasure data, but authentic speech varies with speaker, content, recording environment, and codec, so a compact normal cluster can be an oversimplification.

Self-supervised speech encoders provide a practical starting point. WavLM is pretrained on large speech corpora and exposes representations at multiple depths~\cite{chen2022wavlm}. As a supervised reference we use \supref{}, which replaces the LFCC (Linear Frequency Cepstral Coefficients) input of HM-Conformer~\cite{shin2023hm} with frozen WavLM Base+ features and trains a hierarchical Conformer back-end with multi-level OC-Softmax heads on both bona fide and spoof labels. That design improves the ASVspoof 2021 DF Equal Error Rate (EER) from 15.71\% (LFCC HM-Conformer) to 14.94\% and was trained under conditions that corresponded to this study. It is not a bona-fide-only detector: OC-Softmax uses spoof labels during optimization, so \supref{} cannot answer whether frozen WavLM representations alone support open-world normality modeling.

We ask whether frozen multi-level WavLM representations can support a strict attack-free protocol: no spoof sample is used to fit preprocessing, normality parameters, score normalization, fusion, or the operating threshold. \system{} keeps the recipe simple: each selected layer receives an independent shrinkage-Mahalanobis normality model, and the standardized distances are averaged with equal weights. No task-specific neural back-end is trained. \supref{} remains in the study only as a supervised reference computed from the same frozen WavLM front-end family and the same evaluation partitions.

The study is organized around four research questions:
\begin{itemize}
\item \textbf{RQ1.} Can an attack-free normality model on frozen WavLM layers approach supervised ranking quality in the source domain?
\item \textbf{RQ2.} Do the four encoder depths contribute complementary signal, or is the ensemble dominated by a single layer?
\item \textbf{RQ3.} How do ranking quality and a bona-fide-calibrated threshold behave under the frozen 2021 DF compound domain shift?
\item \textbf{RQ4.} Can the operating threshold be re-estimated from \emph{unlabeled} deployment streams, and how robust is such adaptation to contamination by anomalies and to domain shift affecting bona fide scores?
\end{itemize}

The contributions are:
\begin{itemize}
\item a strict, attack-free protocol that separates bona-fide-only fitting and calibration from labelled comparative evaluation under domain shift;
\item \system{}, a density baseline on frozen WavLM layers with no trained task parameters, and pre-fixed controls for PCA, distance family, KNN, one-class SVM, and multi-centroid scoring;
\item an explicit comparison against a matched supervised \supref{} reference on ASVspoof 2019 LA and ASVspoof 2021 DF, including relative cross-domain degradation, leave-one-layer-out analysis, and transferred-threshold diagnostics;
\item a contamination-controlled study of unlabeled threshold adaptation on simulated deployment streams, quantifying when cheap recalibration restores the nominal operating point and when it self-poisons or merely exposes deeper ranking degradation.
\end{itemize}

\section{Related Work}
\subsection{Supervised audio deepfake countermeasures}
Early spoofing countermeasures paired spectral representations such as LFCC or CQCC with shallow classifiers. End-to-end systems including RawNet2~\cite{tak2021rawnet2} and AASIST~\cite{jung2022aasist} learn discriminative representations from waveforms or spectro-temporal graphs. HM-Conformer aggregates artifacts at multiple temporal resolutions through hierarchical pooling and multi-level classification~\cite{shin2023hm}. These systems are conventionally optimized with both bona fide and spoof examples.

Self-supervised encoders such as wav2vec~2.0, HuBERT, and WavLM have become strong front-ends for spoofing detection~\cite{baevski2020wav2vec,hsu2021hubert,chen2022wavlm,tak2022wav2vec,wang2021investigating}. Full fine-tuning can be effective but is expensive and remains supervised by the attacks represented downstream. WavLM has also been coupled to supervised attentive classifiers~\cite{guo2024wavlm}. The \supref{} reference belongs to this supervised SSL-front-end line: the encoder is frozen, but the HM-Conformer and OC-Softmax heads still consume spoof labels. More recent work selects or fuses SSL layers for generalizable or out-of-domain deepfake detection~\cite{serrano2025layer,beheshti2026probing}. Those papers also use intermediate SSL depths; here the depths stay fixed and are scored without spoof labels. Supervised SSL countermeasures remain the stronger option when labelled attacks are available; the present experiments ask a narrower question about frozen representations under an attack-free constraint.

\subsection{One-class and open-world detection}
OC-Softmax encourages bona fide and spoof embeddings to occupy different angular regions, but its standard anti-spoofing formulation uses both labels during optimization~\cite{zhang2021oneclass}. Recent one-class speech deepfake methods improve centroid estimation and authentic-speech compactness, including Adaptive Centroid Shift and representation-similarity objectives~\cite{kim2024acs,kim2025bpl}. QAMO further models bona fide speech with quality-aware multiple centroids~\cite{truong2025qamo}. More generally, deep one-class classification learns compact normal representations~\cite{ruff2018deep}, while Mahalanobis scores from multiple network layers are established tools for out-of-distribution detection~\cite{lee2018mahalanobis}.

Open-world and continual anomaly detection also note that the normal class can drift after deployment. A useful test is then not only unseen attacks, but whether a threshold fitted on source-domain bona fide scores still holds when authentic conditions change. ASVspoof 2019 LA $\rightarrow$ 2021 DF is a convenient proxy for that compound shift: previously unseen generators, different source material, and lossy compression all move both spoof and bona fide score distributions. Unlike neural one-class deepfake detectors, the models here leave WavLM frozen and do not train a back-end. Equal-weight fusion is also fixed before labelled evaluation, rather than learned from spoof data. This is a statistical baseline on multiple frozen depths. We do not argue that bona fide speech is a single Gaussian, and we do not present the method as a substitute for ACS, QAMO, or a supervised leaderboard system.

\section{Two Systems Under Comparison}
The two WavLM-based systems are easy to mix up, so we keep them apart below.

\paragraph{Proposed: \system{}.} Frozen WavLM Base+ produces pooled embeddings at layers $\ell\in\{3,6,9,12\}$. For each layer we fit StandardScaler, an optional PCA map, and a Ledoit--Wolf Mahalanobis model using only 2019 LA bona fide training utterances. Layer scores are robustly normalized on the same bona fide set and averaged with equal weights. We write \system{} for this method family and distinguish two pre-specified Mahalanobis variants: \syspca{} (128-D PCA) and \sysfull{} (full standardized space). No Conformer, no OC-Softmax, and no spoof label enters fitting or threshold selection.

\paragraph{Reference: supervised \supref{}.} The same WavLM Base+ front-end feeds a trainable HM-Conformer with hierarchical pooling and multi-level classification aggregation~\cite{shin2023hm}. A linear projection maps 768-D WavLM features to the Conformer width, the back-end aggregates multi-resolution tokens and applies multi-level OC-Softmax heads. Training uses ASVspoof 2019 LA with both bona fide and spoof labels, standard waveform augmentations, and checkpoint selection on development EER. The selected checkpoint (epoch 132; about 4.44\% development EER) is the reference used in this paper and reaches 14.94\% EER on ASVspoof 2021 DF. OC-Softmax is often described as one-class, but the loss still uses spoof labels to push fake embeddings away from the bona fide angular region, so the reference is a supervised countermeasure.

The shared frozen WavLM front-end makes the comparison informative: both systems start from related representations, but only \system{} is constrained to attack-free normality modeling. A gain from WavLM$+$HM-Conformer is therefore not evidence for open-world one-class detection. The two numbers that matter for that contrast are source-domain ranking, where the systems are close, and frozen 2021 DF transfer, where supervision still helps and a source-calibrated threshold does not.

\section{Method: Attack-Free \system{}}
\subsection{Why not train HM-Conformer as one-class?}
A natural alternative would be to keep the full WavLM$+$HM-Conformer stack and remove spoof terms from OC-Softmax, training the back-end only on bona fide speech. We do not do that here. Without a one-class regularizer, a high-capacity Conformer can collapse~\cite{ruff2018deep}. The same experiment would also mix three effects that we want to keep separate: frozen SSL features, trainable hierarchical pooling, and the training objective. Fitting only shallow models on frozen WavLM isolates the first of those. If those models already fail to transfer a threshold under domain shift, the failure is easier to read; it is not an accident of an under-regularized neural back-end.

\subsection{Frozen multi-level representations}
For waveform $x$, frozen WavLM produces hidden sequences $H_\ell(x)\in\mathbb{R}^{T_\ell\times D}$ at transformer depth $\ell$. Padding frames are excluded using the feature-extractor output lengths. An utterance representation is obtained by masked temporal mean pooling:
\begin{equation}
 e_\ell(x)=\frac{1}{T_\ell^\prime}\sum_{t=1}^{T_\ell^\prime}H_{\ell,t}(x).
\end{equation}
We use layers $\ell\in\{3,6,9,12\}$ of WavLM Base+. This set and all subsequent hyperparameters are specified without spoof labels and frozen before labelled evaluation, including evaluation on ASVspoof 2021 DF.

\subsection{Bona-fide-only normality model}
For each layer, a standardizer is fitted on bona fide training embeddings. Two Mahalanobis preprocessing variants were pre-specified before labelled evaluation: a 128-dimensional PCA projection (\syspca{}) and the full standardized space (\sysfull{}). Both variants remain in the comparison; \sysfull{} is not promoted after seeing labelled scores. Let $z_\ell(x)$ denote the transformed representation. We estimate its location $\mu_\ell$ and a Ledoit--Wolf shrinkage covariance $\Sigma_\ell$~\cite{ledoit2004well}. The layer anomaly score is
\begin{equation}
 d_\ell(x)=\sqrt{(z_\ell(x)-\mu_\ell)^\top\Sigma_\ell^{-1}(z_\ell(x)-\mu_\ell)}.
\end{equation}
We also evaluate the Euclidean distance to the bona fide centroid after the same preprocessing. Distances differ in scale across encoder depths. For each layer we therefore estimate the median $m_\ell$ and robust scale $s_\ell=1.4826\,\mathrm{MAD}(d_\ell)$ on the bona fide training data:
\begin{equation}
 \widetilde d_\ell(x)=\frac{d_\ell(x)-m_\ell}{\max(s_\ell,\epsilon)}.
\end{equation}
The final score contains no learned fusion parameters:
\begin{equation}
 S(x)=\frac{1}{4}\sum_{\ell\in\{3,6,9,12\}}\widetilde d_\ell(x).
\end{equation}
Higher values indicate a greater deviation from the authentic-speech model.

\subsection{Pre-fixed alternative scorers}
To avoid evaluating only variants of one estimator, we include three additional bona-fide-only scorers with hyperparameters fixed before evaluation:
\begin{itemize}
\item KNN distance with $k=5$ after PCA-128;
\item RBF one-class SVM with $\nu=0.05$ and $\gamma=\texttt{scale}$ after PCA-128;
\item multi-centroid distance with $k=8$ $k$-means centroids fitted on bona fide data only; the score is the Euclidean distance to the nearest centroid, with and without PCA.
\end{itemize}
None of these controls uses spoof labels for model selection.

\subsection{Calibration and decision rule}
Ranking metrics are evaluated directly from $S(x)$. For an operational one-class threshold, the 99th percentile is estimated using only bona fide development scores. We separately report this operating point because EER and ROC-AUC use labelled evaluation data to characterize ranking and do not constitute a deployable bona-fide-only threshold. As a post-hoc diagnostic, we also transfer a mixed-development EER threshold that uses spoof labels; this diagnostic is not part of the attack-free system.

\subsection{Unlabeled threshold adaptation}\label{sec:adaptation-method}
Open-world deployments observe unlabeled target-domain audio, so the cheapest plausible adaptation is to re-estimate the operating threshold from the incoming stream without labels. We compare three strategies against the frozen source threshold:
\begin{itemize}
\item \emph{Stream quantile}: the 99th percentile of the unlabeled stream scores, valid only if the stream is almost entirely bona fide;
\item \emph{Robust median/MAD transfer}: the source calibration is stored as an offset in robust $z$-units, $z^\ast=(q_{99}-m)/(1.4826\,\mathrm{MAD})$ computed on development bona fide scores, and re-anchored on the stream as $\tau=m_{\mathrm{stream}}+z^\ast\cdot1.4826\,\mathrm{MAD}_{\mathrm{stream}}$; median and MAD tolerate a minority of contaminating anomalies;
\item \emph{Oracle}: the 99th percentile of labelled target bona fide scores, reported only as a diagnostic upper bound.
\end{itemize}
Streams are simulated from frozen evaluation scores: half of the bona fide trials form the adaptation stream together with spoof trials sampled to a controlled prevalence $\pi\in\{0,1,2,5,10,25,50\%,\text{natural}\}$; FPR is then measured on the held-out bona fide half and TPR on spoof trials not used for adaptation, over ten pre-fixed seeds. The prevalence grid, seeds, and strategies were fixed before inspecting any result. Labels are used by the experimenter to construct prevalence-controlled streams and to evaluate FPR/TPR; the adaptation algorithm itself receives only unlabeled scores, and no 2021 label influences the adaptive thresholds.

\subsection{Protocol checklist}
The attack-free pipeline is summarized as follows:
\begin{enumerate}
\item Extract frozen WavLM embeddings for layers $\{3,6,9,12\}$ with masked mean pooling; cache complete utterance embeddings.
\item Fit per-layer scaler / optional PCA / Mahalanobis (or control scorer) on the 2{,}580 bona fide training utterances only.
\item Normalize each layer score with train bona fide median/MAD; form the equal-weight ensemble.
\item Set the operating threshold to the 99th percentile of development bona fide ensemble scores.
\item Evaluate ranking and fixed-threshold metrics on 2019 LA evaluation and, without any further selection, on 2021 DF.
\item Simulate unlabeled deployment streams at pre-fixed contamination levels and re-estimate thresholds per Section~\ref{sec:adaptation-method}.
\end{enumerate}
Spoof labels appear in step~5, in the optional mixed-threshold diagnostic, in the oracle row of step~6, and in the experimenter's construction of prevalence-controlled streams; the adaptation algorithm itself never sees labels.

\section{Experimental Protocol}
\subsection{Datasets and open-world split}
ASVspoof 2019 LA provides the source domain~\cite{todisco2019asvspoof}. Only its 2,580 bona fide training utterances are used to fit the proposed normality models. The development partition contains 2,548 bona fide and 22,296 spoofed utterances; spoof labels are used only to report development metrics and the post-hoc mixed-threshold diagnostic, while the operational threshold uses the bona fide subset. The 2019 LA evaluation partition contains 7,355 bona fide and 63,882 spoofed utterances.

ASVspoof 2021 DF is the frozen cross-domain test set~\cite{liu2023asvspoof,yamagishi2021asvspoof}. It introduces different synthesis systems and codec conditions and was not used to select layers, pooling, PCA dimension, covariance estimator, score normalization, fusion, or thresholds. Of 611,829 trials, 611,828 were decoded successfully. One corrupted file rejected by two independent audio decoders was excluded and logged by sample identifier; no zero waveform was substituted. Exact sample-ID checks across train, development, and evaluation caches found no overlap. ASVspoof~5 provides a newer evaluation context but is not used in the present fit or model-selection protocol~\cite{wang2026asvspoof5}.

\subsection{Supervised reference details}
For completeness we briefly record the supervised reference configuration used for comparison. \supref{} freezes WavLM Base+, projects features into an HM-Conformer stack with hierarchical pooling and multi-level classification aggregation, and optimizes multi-level OC-Softmax with both class labels. Inputs are 4\,s mono waveforms at 16~kHz. The checkpoint selected on 2019 LA development EER (epoch 132) is evaluated unchanged on 2019 LA evaluation and 2021 DF; no 2021 label is used for checkpoint choice. This isolates the effect of allowing spoof supervision, rather than changing the front-end family.

\subsection{Implementation and compute}
All audio is converted to mono at 16~kHz and center-cropped or zero-padded to four seconds. WavLM Base+ is frozen and evaluated without stochastic augmentation. Padding is excluded during temporal pooling. Embedding extraction runs on a single consumer GPU with 12 GB VRAM. The number of gradient-optimized parameters in \system{} is zero.

\begin{table}[t]
\centering
\caption{Frozen embedding caches used for all one-class scoring. Shapes are rows $\times$ layers $\times$ embedding dimension.}
\label{tab:compute}
\begin{tabular}{lrr}
\toprule
Artifact & Shape / dtype & Size\\
\midrule
Train bona fide cache & $2580\times4\times768$ f32 & 30.5\,MB\\
Dev cache & $24844\times4\times768$ f32 & 294\,MB\\
2019 LA eval cache & $71237\times4\times768$ f32 & 843\,MB\\
2021 DF eval cache & $611828\times4\times768$ f32 & 7.07\,GB\\
2021 DF score cache & $611828\times6\times5$ f32 & 103.5\,MB\\
Frozen WavLM Base+ & \multicolumn{2}{r}{$\sim$94.7\,M params (not trained)}\\
Trainable params (\system{}) & 0 & ---\\
\bottomrule
\end{tabular}
\end{table}

\subsection{Metrics and uncertainty}
We report EER, ROC-AUC, TPR at 1\% and 5\% FPR, and fixed-threshold FPR/TPR at the bona-fide-calibrated operating point. Given the frozen embedding caches and fixed seed, every one-class fit is deterministic, so run-to-run variance is zero by construction; uncertainty is therefore quantified by resampling. Bootstrap confidence intervals use 2{,}000 class-stratified replicates with seed 42; on 2021 DF we additionally use system/codec cluster bootstrap, and the adaptation study aggregates ten stream-sampling seeds. Paired comparisons reuse the same resampled trials for both systems. Per-codec and per-system analyses use official metadata where available. The community min-TDCF metric requires ASV scores that are not part of this protocol; a false-alarm-averse detection cost ($P_{\mathrm{tar}}{=}0.05$, $C_{\mathrm{miss}}{=}1$, $C_{\mathrm{fa}}{=}10$) gives normalized minDCF 0.90 on 2019 LA and $\approx$1.0 on 2021 DF.

\section{Results}
\subsection{Source-domain one-class ranking (RQ1)}
Table~\ref{tab:source} reports the strict one-class evaluation on ASVspoof 2019 LA. Mahalanobis improves substantially over centroid, KNN, OCSVM, and multi-centroid scoring. Contrary to the initial design assumption, retaining all standardized dimensions improves every centroid and Mahalanobis configuration.

\begin{table}[t]
\centering
\caption{ASVspoof 2019 LA evaluation. All rows except the supervised reference fit bona fide only. Ensemble: equal mean over layers 3/6/9/12. \syspca{} and \sysfull{} are the pre-specified Mahalanobis variants.}
\label{tab:source}
\begin{tabular}{lrrr}
\toprule
System & EER (\%) & AUC & TPR@1\\
~ & ~ & ~ & \%FPR\\
\midrule
Centroid, PCA-128 & 19.05 & 0.866 & 7.95\\
\syspca{} & 12.40 & 0.938 & 43.69\\
Centroid, no PCA & 16.52 & 0.895 & 12.86\\
\textbf{\sysfull{}} & \textbf{8.45} & \textbf{0.966} & \textbf{69.17}\\
$k$NN ($k=5$), PCA-128 & 21.35 & 0.871 & 20.92\\
RBF OCSVM, PCA-128 & 15.52 & 0.908 & 23.03\\
Multi-centroid ($k=8$), PCA-128 & 20.76 & 0.867 & ---\\
Multi-centroid ($k=8$), no PCA & 15.73 & 0.911 & ---\\
\midrule
Supervised \supref{} & 7.86 & 0.941 & 79.42\\
\bottomrule
\end{tabular}
\end{table}

\begin{figure}[t]
\centering
\includegraphics[width=\columnwidth]{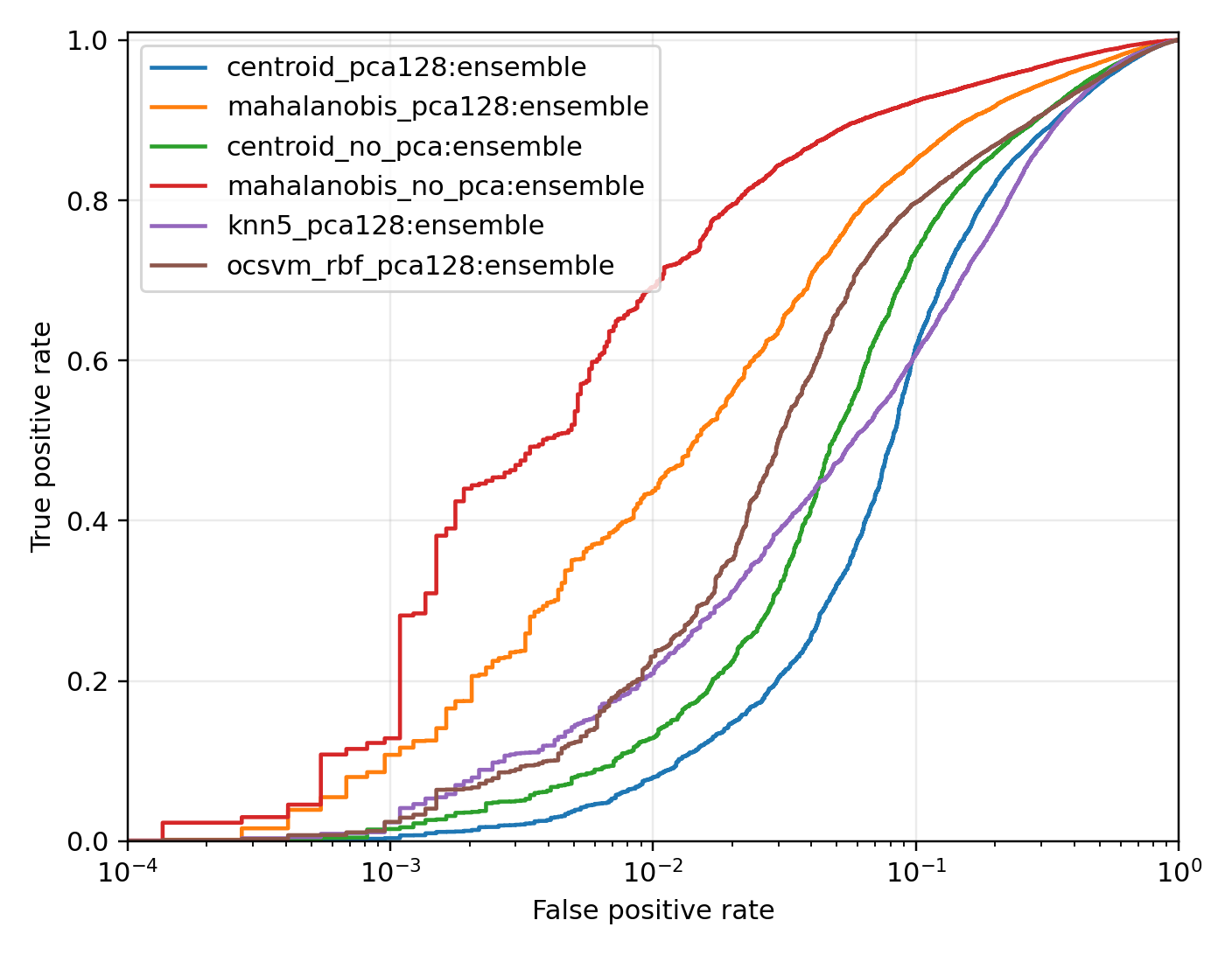}
\caption{ROC curves on ASVspoof 2019 LA evaluation for selected bona-fide-only systems. Shrinkage Mahalanobis without PCA dominates the other one-class controls.}
\label{fig:la-roc}
\end{figure}

The pre-specified \sysfull{} ensemble improves from 12.40\% to 8.45\% EER relative to \syspca{}. At 1\% and 5\% FPR it detects 69.17\% and 88.68\% of spoof trials. Its 2{,}000-replicate EER interval is $[8.18,8.72]\%$. Paired bootstrap on 2019 LA finds that \syspca{} is worse than \sysfull{} by about $+3.96$ EER points (95\% CI excludes zero). A pre-fixed $k=8$ multi-centroid baseline remains weaker than shrinkage Mahalanobis, indicating that a single regularized Gaussian already captures the dominant bona fide geometry on this source set. The supervised \supref{} reference obtains 7.86\% EER, 0.941 ROC-AUC, and 79.42\% TPR at 1\% FPR on the same 2019 evaluation partition: better in EER and low-FPR recall than the best one-class model, and slightly worse in AUC (0.941 vs. 0.966). At 5\% FPR the two systems are close (88.98\% supervised vs. 88.68\% \sysfull{}). Figure~\ref{fig:la-roc} shows that the best one-class ROC already separates classes strongly in-domain. The remaining supervised gain is concentrated near very low FPR.

\begin{figure}[t]
\centering
\includegraphics[width=\columnwidth]{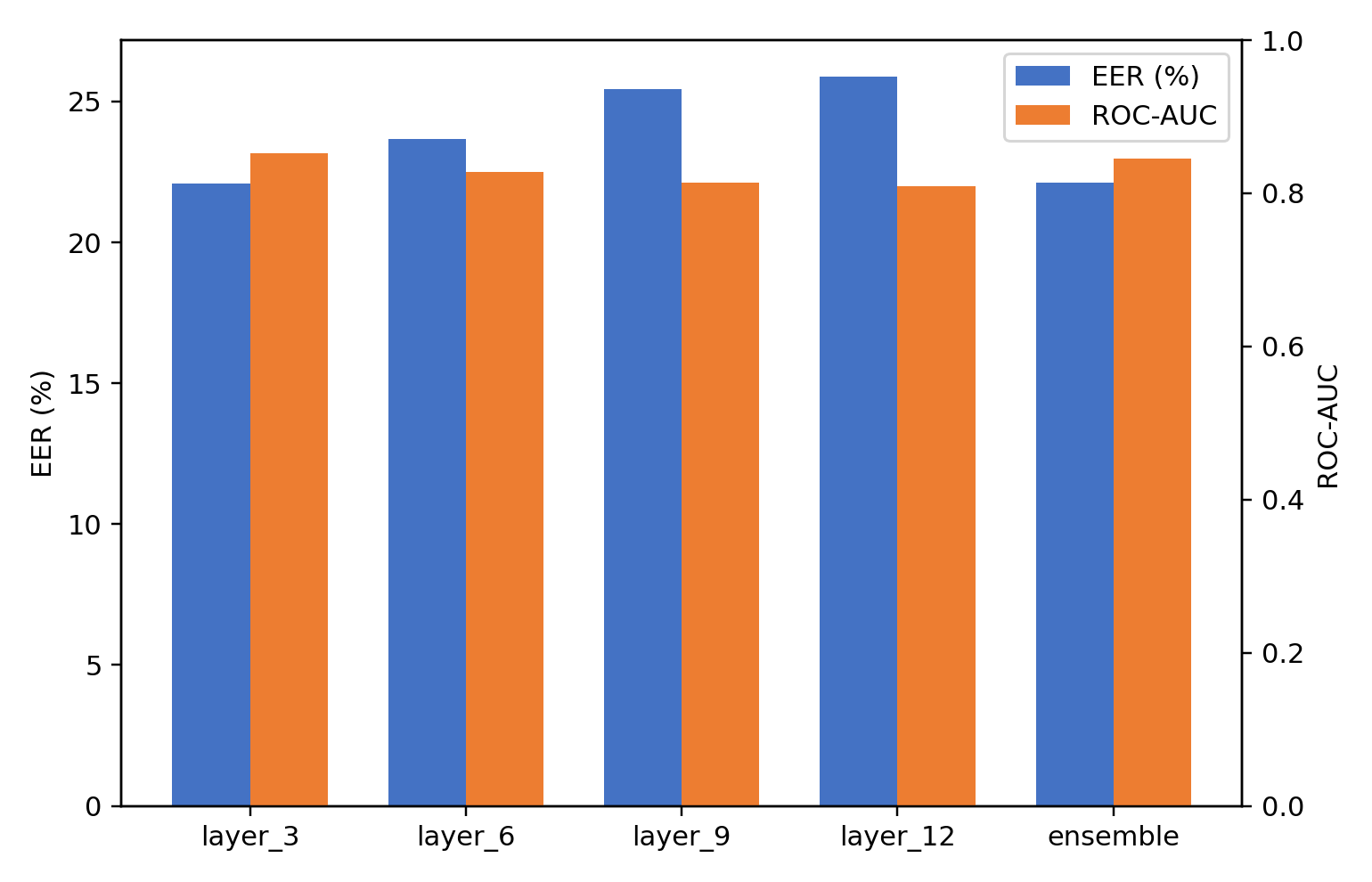}
\caption{ASVspoof 2019 LA: layer-wise and ensemble metrics for the Mahalanobis configurations. Early layers are strong; equal fusion helps most relative to later layers.}
\label{fig:la-layers}
\end{figure}

Table~\ref{tab:layers} reports per-layer EER for both Mahalanobis settings. For PCA-128, the equal ensemble improves over every individual layer on both partitions. For no-PCA, the ensemble still helps on 2019 LA, but on 2021 DF it is essentially tied with layer~3. In short, multi-layer averaging helps as a safeguard, but the cross-domain signal is concentrated early in the encoder.

\begin{table}[t]
\centering
\caption{Per-layer EER (\%) for shrinkage Mahalanobis. Ensemble uses equal weights fixed before evaluation.}
\label{tab:layers}
\begin{tabular}{lrrrr}
\toprule
 & \multicolumn{2}{c}{PCA-128} & \multicolumn{2}{c}{No PCA}\\
\cmidrule(lr){2-3}\cmidrule(lr){4-5}
Layer & 2019 & 2021 & 2019 & 2021\\
\midrule
3 & 14.52 & 25.48 & 9.22 & 22.10\\
6 & 19.01 & 29.14 & 11.70 & 23.68\\
9 & 17.46 & 28.32 & 12.77 & 25.43\\
12 & 17.50 & 27.07 & 14.11 & 25.88\\
Ensemble & 12.40 & 25.06 & 8.45 & 22.12\\
\bottomrule
\end{tabular}
\end{table}

\subsection{Leave-one-layer-out and score correlation (RQ2)}
Table~\ref{tab:loo} summarizes leave-one-layer-out on the frozen \sysfull{} scores. Layer~3 carries most of the gain: removing it raises the 2019 ensemble from 8.45\% to 11.64\% EER, whereas removing layer~6, 9, or 12 changes EER by less than 0.3 points and can slightly help. Pearson correlations among the four layer scores are high (mean $|\rho|=0.73$ on 2019 LA; $0.78$ on 2021 DF), so the depths are only partly complementary. Equal-weight fusion is therefore a pre-fixed rule, not evidence that the four depths specialize.

\begin{table}[t]
\centering
\caption{Leave-one-layer-out for the \sysfull{} ensemble. Positive $\Delta$EER means removing the layer hurts the ensemble.}
\label{tab:loo}
\begin{tabular}{lrrrr}
\toprule
 & \multicolumn{2}{c}{2019 LA} & \multicolumn{2}{c}{2021 DF}\\
\cmidrule(lr){2-3}\cmidrule(lr){4-5}
Score & EER & $\Delta$EER & EER & $\Delta$EER\\
\midrule
Full ensemble & 8.45 & --- & 22.12 & ---\\
Drop layer 3 & 11.64 & $+3.19$ & 24.19 & $+2.07$\\
Drop layer 6 & 8.17 & $-0.28$ & 22.20 & $+0.08$\\
Drop layer 9 & 8.40 & $-0.05$ & 21.84 & $-0.28$\\
Drop layer 12 & 8.28 & $-0.16$ & 21.52 & $-0.60$\\
Layer 3 alone & 9.22 & --- & 22.10 & ---\\
\bottomrule
\end{tabular}
\end{table}

\subsection{Frozen cross-domain evaluation (RQ3)}
Table~\ref{tab:crossdomain} reports ASVspoof 2021 DF. \sysfull{} remains the strongest bona-fide-only system, but absolute performance is substantially weaker than on 2019 LA.

\begin{table}[t]
\centering
\caption{Frozen ASVspoof 2021 DF results. The supervised reference uses spoof labels during training; all other rows are bona-fide-only.}
\label{tab:crossdomain}
\begin{tabular}{lrrr}
\toprule
System & EER (\%) & AUC & TPR@5\\
~ & ~ & ~ & \%FPR (\%)\\
\midrule
Centroid, PCA-128 & 29.36 & 0.764 & 6.57\\
\syspca{} & 25.06 & 0.807 & 8.58\\
Centroid, no PCA & 26.80 & 0.792 & 7.63\\
\textbf{\sysfull{}} & \textbf{22.12} & \textbf{0.845} & \textbf{15.20}\\
KNN ($k=5$), PCA-128 & 28.65 & 0.771 & 9.82\\
RBF OCSVM, PCA-128 & 26.13 & 0.799 & 8.10\\
Multi-centroid ($k=8$), PCA-128 & 29.26 & 0.765 & ---\\
Multi-centroid ($k=8$), no PCA & 25.62 & 0.800 & ---\\
\midrule
Supervised \supref{} & 14.94 & 0.914 & 58.63\\
\bottomrule
\end{tabular}
\end{table}

\begin{figure}[t]
\centering
\includegraphics[width=\columnwidth]{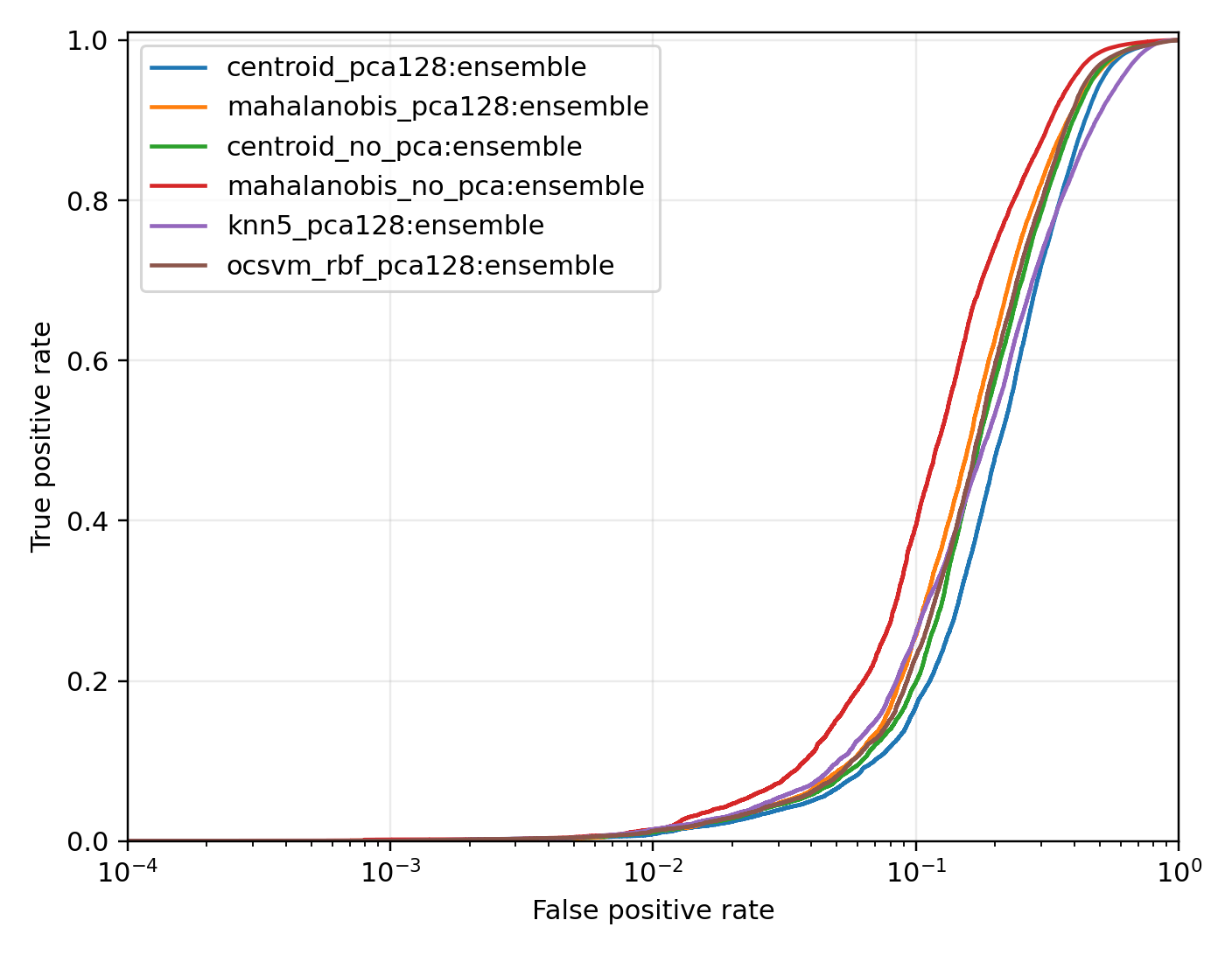}
\caption{ROC curves on frozen ASVspoof 2021 DF. Ranking remains informative, but the operating region near low FPR is weak.}
\label{fig:df-roc}
\end{figure}

Relative to the same supervised checkpoint on 2019 LA evaluation (7.86\%~$\rightarrow$~14.94\% EER; $+7.08$ points), the best one-class variant degrades more (8.45\%~$\rightarrow$~22.12\%; $+13.67$ points). The attack-free detector therefore has a larger cross-domain EER drop, not a smaller one (Table~\ref{tab:relative}). The gap is wider at low FPR: at 5\% FPR, supervised TPR falls from 88.98\% to 58.63\%, whereas \sysfull{} falls from 88.68\% to 15.20\%; at 1\% FPR the corresponding figures are 79.42\%~$\rightarrow$~15.98\% versus 69.17\%~$\rightarrow$~1.46\%. That is more than a shifted threshold: ranking itself is worse, including relative to the supervised system at a 1\% false-alarm budget. On 2021 DF, leave-one-layer-out again identifies layer~3 as critical, while the full \sysfull{} ensemble (22.12\%) nearly ties layer~3 alone (22.10\%). The supervised result was independently reproduced from saved per-trial scores: 14.941\% EER and 0.9138 ROC-AUC over all 611{,}829 trials. Excluding the single spoof trial that one-class extraction could not decode leaves the supervised EER unchanged at 14.94\%. The gap to \syspca{} remains 10.12 points (paired bootstrap advantage of supervised over \syspca{}: 10.11 points, 95\% CI $[9.79,10.44]$).

\begin{table}[t]
\centering
\caption{Relative cross-domain degradation for the same supervised checkpoint and the two pre-specified Mahalanobis variants.}
\label{tab:relative}
\begin{tabular}{lrrr}
\toprule
System & 2019EER & 2021EER & $\Delta$EER\\
\midrule
Supervised \supref{} & 7.86 & 14.94 & $+7.08$\\
\sysfull{} & 8.45 & 22.12 & $+13.67$\\
\syspca{} & 12.40 & 25.06 & $+12.66$\\
\bottomrule
\end{tabular}
\end{table}

\subsection{Operating points and codec shift}
The low-FPR regime remains challenging: \sysfull{} detects only 1.46\% of spoof trials at 1\% FPR and 15.20\% at 5\% FPR on 2021 DF, versus 15.98\% and 58.63\% for the supervised reference. Applying its 99th-percentile bona fide threshold calibrated on 2019 development data yields 20.23\% false positives and 74.81\% true positives on 2021 DF (Table~\ref{tab:threshold}). A post-hoc mixed-development EER threshold raises recall to 98.44\%, but also inflates the false-positive rate to 49.86\% under the same shift. The spoof-aware source-domain calibration therefore does not repair the moving bona fide reference.

\begin{table}[t]
\centering
\caption{Fixed-threshold transfer for the \sysfull{} ensemble. The mixed-dev EER threshold is a post-hoc diagnostic only.}
\label{tab:threshold}
\begin{tabular}{lrrrr}
\toprule
Threshold source & \multicolumn{2}{c}{2019 LA} & \multicolumn{2}{c}{2021 DF}\\
\cmidrule(lr){2-3}\cmidrule(lr){4-5}
 & FPR & TPR & FPR & TPR\\
\midrule
Bona fide q99 (attack-free) & 0.61\% & 60.07\% & 20.23\% & 74.81\%\\
Mixed-dev EER (diagnostic) & 9.10\% & 91.92\% & 49.86\% & 98.44\%\\
\bottomrule
\end{tabular}
\end{table}

At the bona-fide-calibrated operating point, per-codec false-positive rates on 2021 DF bona fide utterances range from 18.50\% (high-quality OGG) to 25.31\% (MP3--M4A), with elevated errors for low-bitrate MP3 (23.84\%). Codec-conditioned ranking EER for the same model spans roughly 20--23\% across codec groups, a narrower range than the variation across synthesis systems. Using the complete bona fide set as a common reference, per-system EER spans 10.70\% to 51.74\% across 110 labelled systems. The hardest system (A19) has ROC-AUC below 0.5, indicating a reversed score ordering rather than merely poor threshold calibration. This spread is a reason to report attack-level numbers, not only the aggregate EER.

\begin{figure}[t]
\centering
\includegraphics[width=\columnwidth]{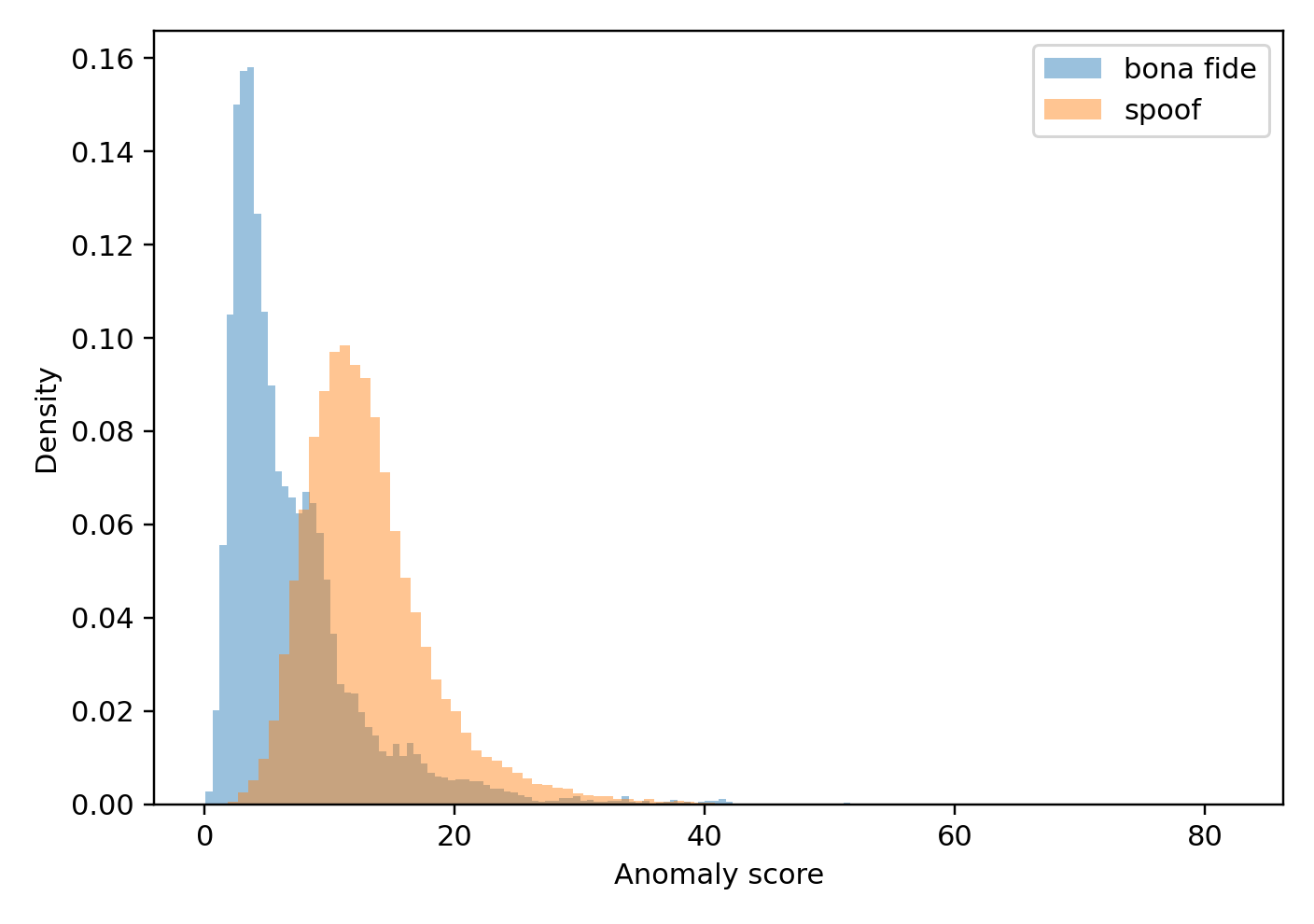}
\caption{Score distributions on 2021 DF dataset. Overlap between bona fide and spoof scores explains why a source-calibrated threshold transfers poorly.}
\label{fig:df-scores}
\end{figure}

\subsection{Unlabeled threshold adaptation under contamination (RQ4)}
\begin{figure}[t]
\centering
\includegraphics[width=\columnwidth]{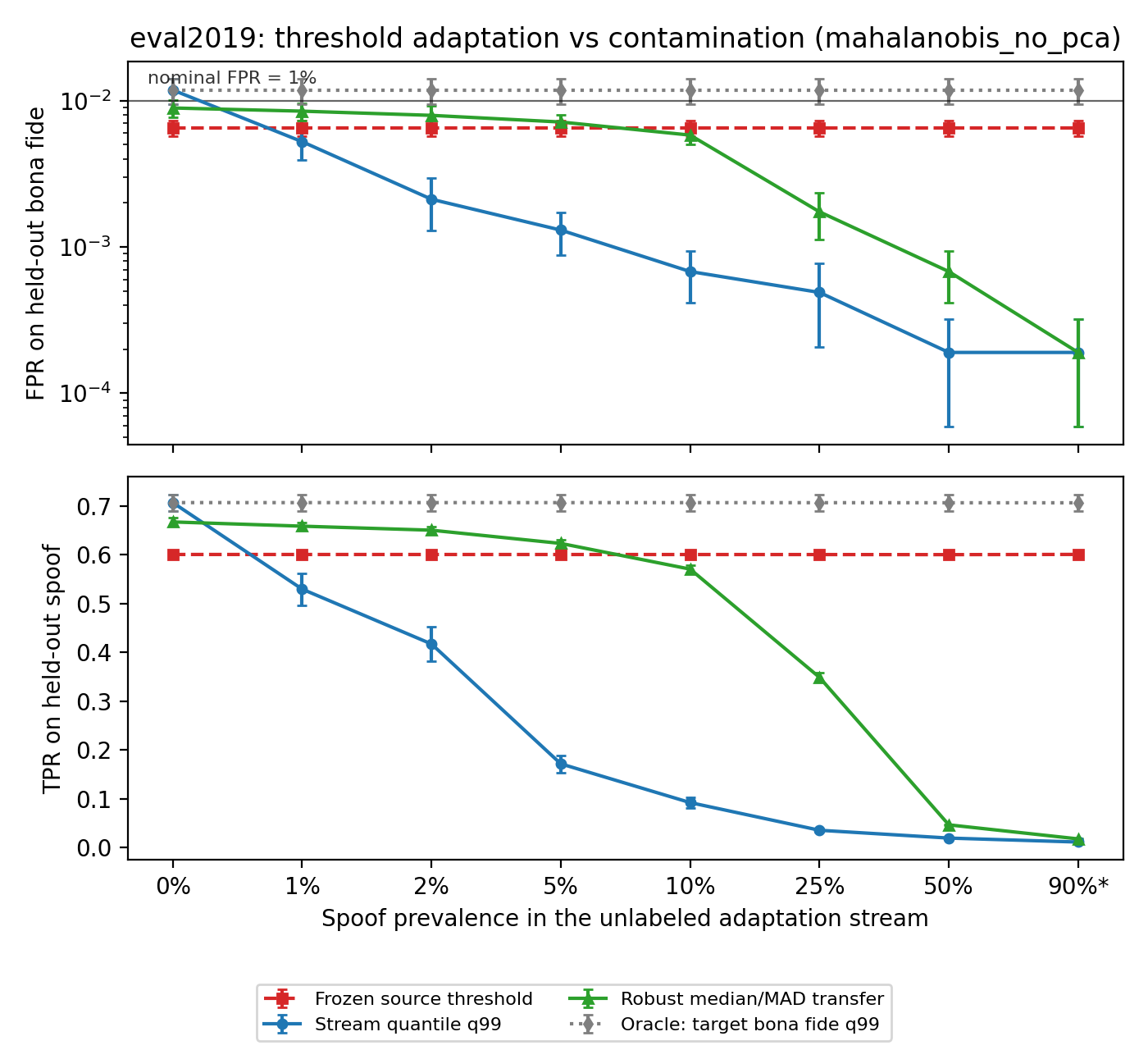}
\caption{2019 LA dataset: unlabeled threshold adaptation versus stream contamination (\sysfull{}, mean $\pm$ std over 10 seeds). Robust MAD transfer keeps recall up to about 10\% spoof prevalence; the naive stream quantile does not.}
\label{fig:adaptation-la}
\end{figure}

Table~\ref{tab:adaptation2019} reports in-domain adaptation on 2019 LA evaluation streams for the \sysfull{} ensemble. On a clean stream, unlabeled quantile recalibration exactly recovers the oracle operating point (1.2\% FPR, 70.7\% TPR), improving on the slightly conservative frozen threshold (0.7\% FPR, 60.1\% TPR). The two unlabeled rules then diverge. The naive stream quantile self-poisons quickly: 5\% spoof prevalence already drags recall from 70.7\% to 17.2\% because contaminating anomalies inflate the estimated 99th percentile. The robust median/MAD transfer falls off more slowly, retaining 62.3\% TPR at 5\% and 57.1\% at 10\% contamination, and only collapses once anomalies stop being a minority (25\% and above). In-domain, unlabeled adaptation is usable while spoofs remain rare, if the threshold is estimated with robust statistics. Figure~\ref{fig:adaptation-la} shows the same split: the naive quantile matches the oracle only on a clean stream, while robust MAD transfer stays near the 1\% false-alarm budget until contamination exceeds about 10\%.

\begin{table}[t]
\centering
\caption{Unlabeled threshold adaptation on 2019 LA streams (\sysfull{}, mean over 10 seeds). Frozen threshold: 0.65\% FPR / 60.1\% TPR. Oracle: 1.18\% FPR / 70.7\% TPR at every prevalence.}
\label{tab:adaptation2019}
\begin{tabular}{lrrrr}
\toprule
Spoof & \multicolumn{2}{c}{Stream quantile} & \multicolumn{2}{c}{Robust transfer}\\
\cmidrule(lr){2-3}\cmidrule(lr){4-5}
prevalence & FPR & TPR & FPR & TPR\\
\midrule
0\% & 1.18\% & 70.7\% & 0.89\% & 66.7\%\\
1\% & 0.52\% & 53.0\% & 0.85\% & 65.9\%\\
5\% & 0.13\% & 17.2\% & 0.72\% & 62.3\%\\
10\% & 0.07\% & 9.2\% & 0.58\% & 57.1\%\\
25\% & 0.05\% & 3.6\% & 0.17\% & 34.9\%\\
50\% & 0.02\% & 2.0\% & 0.07\% & 4.7\%\\
\bottomrule
\end{tabular}
\end{table}

\begin{figure}[t]
\centering
\includegraphics[width=\columnwidth]{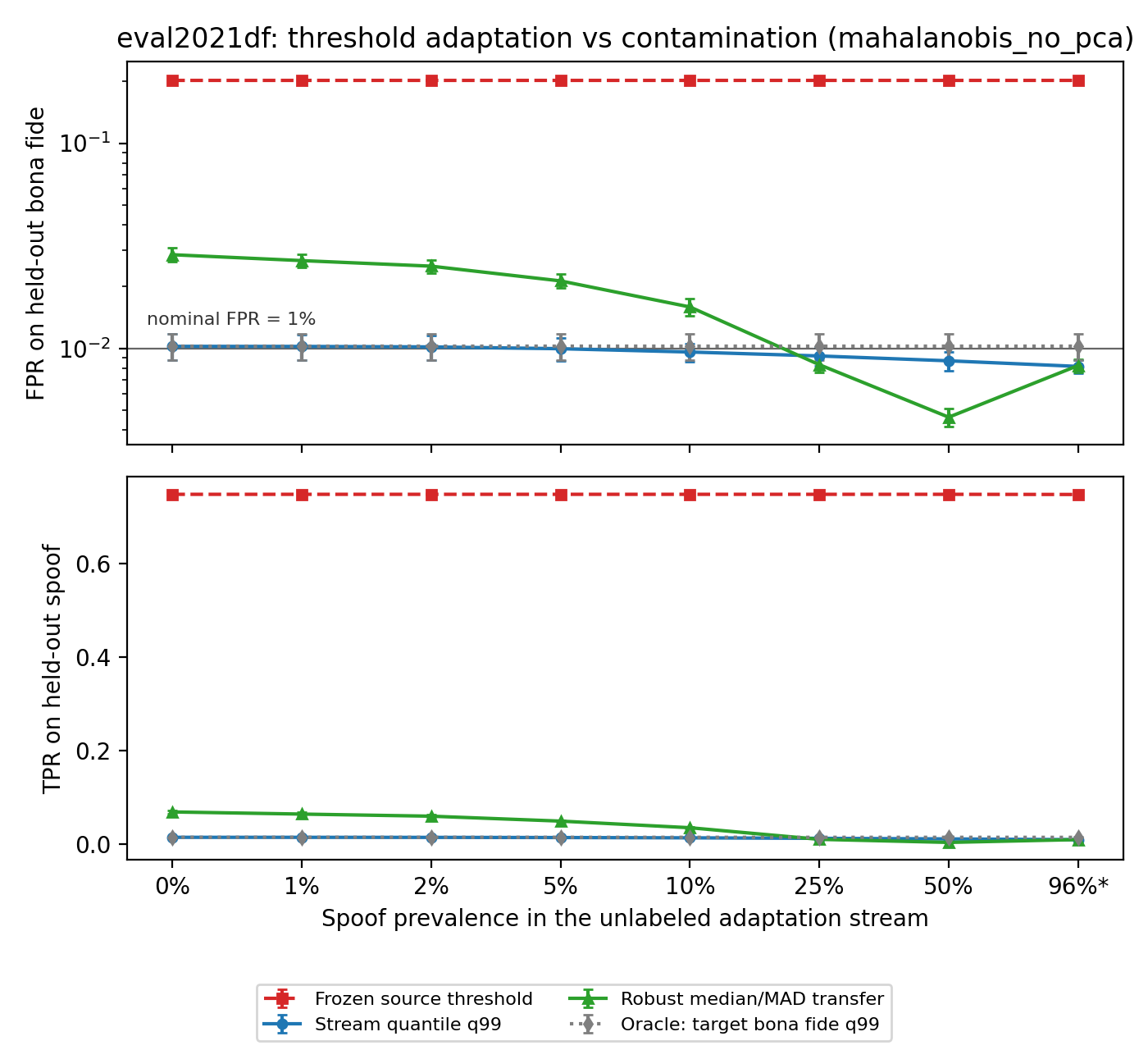}
\caption{2021 DF dataset: threshold adaptation versus stream contamination (Mahalanobis \sysfull{}, mean $\pm$ std over 10 seeds; $96\%^{*}$ is the natural eval prevalence). Adaptive thresholds restore the nominal FPR (top) but cannot restore recall (bottom): at a 1\% false-alarm budget the shifted-domain ranking supports only $\approx$1.5\% TPR even with oracle calibration.}
\label{fig:adaptation-df}
\end{figure}

Under the 2021 DF shift, the outcome changes qualitatively (Figure~\ref{fig:adaptation-df}). All adaptive strategies again restore the false-positive budget: the stream quantile holds FPR near the nominal 1\% at every contamination level, versus 20.2\% for the frozen threshold. But recall at that budget collapses to about 1.5\%, for the unlabeled strategies and for the oracle, matching the 1.46\% TPR at 1\% FPR on the ROC. The frozen threshold still catches 74.8\% of spoofs, but only at 20.2\% FPR, about twenty times the nominal budget. Resetting the threshold recovers the false-positive rate and shows the remaining problem: the low-FPR part of the ranking has already collapsed, so no scalar cut-off restores recall. Adaptation then has to update the bona fide model, not only $\tau$.

\section{Discussion}
\paragraph{Relation to \supref{}} \system{} is a bona-fide-only Mahalanobis ensemble on frozen WavLM, with the pre-specified variants \syspca{} and \sysfull{}. \supref{} uses the same front-end with a trainable HM-Conformer and spoof labels. Close in-domain ranking numbers do not make the two systems the same method; the reference mainly shows how far this front-end family can go when spoof supervision is allowed.

\paragraph{Layer fusion} An early layer is individually strongest, layer-score correlations are high, and leave-one-layer-out shows that later layers contribute little once layer~3 is present. On 2021 DF the no-PCA ensemble nearly ties layer~3 alone. Equal fusion is still a reasonable default so that one does not commit to a late layer in advance, but the four depths are not strongly specialized.

\paragraph{Threshold transfer} Fitting only on 2019 bona fide speech does not give a smaller EER drop than supervised detection ($+13.67$ vs. $+7.08$ points). Removing PCA improves ranking, but bona fide scores also move: operational FPR grows from about 1\% on 2019 development to 20.23\% on 2021 DF. Many spoofs still rank above bona fide trials, yet a threshold fixed before deployment is not usable. ASVspoof 2021 DF mixes generators, source data, and codecs, so we cannot split WavLM feature shift from covariance misfit. The scores themselves have moved, not only the cut-off.

The score is a distance from the 2019 bona fide fit. Channel, speaker, language, and recording changes can raise it as well, which is why per-codec false-positive rates matter as much as spoof recall.

\paragraph{Unlabeled recalibration} If the ranking is still informative, robust MAD transfer recovers a 1\% false-alarm budget up to about 10\% contamination; a naive stream quantile does not, because even a few percent of spoofs inflate the estimated 99th percentile. If the ranking has already collapsed at low FPR, restoring that budget cannot restore recall: every strategy, including the oracle target-bona-fide quantile, meets the same 1.5\% TPR ceiling on 2021 DF. Reporting only the frozen-threshold TPR (74.8\% at 20.2\% FPR) hides that distinction. Checking the assumed contamination rate is part of using the unlabeled rule in practice.

\paragraph{Implications} The difficulty is not only unseen generators. A transferred normal class can also fail on codecs, channels, and source-data mismatch that remain authentic speech. Methods that update the bona fide model online, or that treat codec and channel as nuisance factors, address this more directly than swapping the distance or the threshold. In-domain, \sysfull{} and \supref{} are close. The gap appears under shift and at a fixed operating point.

The experiments support an attack-free protocol, a strong statistical baseline on frozen WavLM, a negative result on transferred thresholds, and a contamination-controlled look at unlabeled adaptation. They do not show that \system{} replaces \supref{}, that the four layers are strongly complementary, or that ASVspoof 2021 DF covers deployed open-world audio.

\section{Threats to Validity}
\paragraph{Construct validity} EER and ROC-AUC describe ranking. Deployability is reported separately as FPR/TPR at a bona-fide-calibrated threshold. A high \system{} score is a distance from 2019 LA bona fide embeddings, not a proof that the clip is synthetic.

\paragraph{Internal validity} All one-class hyperparameters were fixed before 2021 evaluation, and sample-ID checks found no train/dev/eval overlap. Equal-weight fusion and the layer set were chosen from prior SSL practice, not 2021 labels; leave-one-layer-out exposes that sensitivity.

\paragraph{External validity} ASVspoof 2021 DF is a strong but curated compound shift and may not transfer to ASVspoof~5, multilingual speech, or other codec stacks. The supervised reference is one checkpoint, not an exhaustive leaderboard.

\paragraph{Comparison fairness} \supref{} receives spoof supervision and a trainable back-end; \system{} does not. The comparison is the same front-end family with and without spoof labels.

\section{Limitations}
The study uses one English-language benchmark family. WavLM inference remains expensive despite zero task-specific neural training. Temporal mean pooling does not localize manipulations, and the fixed 4\,s center crop may omit artifacts outside the analyzed fragment. The Gaussian layer models are unimodal approximations to heterogeneous authentic speech. We do not reimplement ACS, QAMO, alternative SSL encoders, or learned layer weighting as numerical competitors; learned weighting would also require a label-free objective to remain attack-free. ASVspoof 2021 DF is a controlled compound shift rather than the full diversity of deployed open-world audio.

\section{Conclusion}
We presented \system{}, a bona-fide-only detector on frozen WavLM layers, and compared it with a matched supervised \supref{} reference. Without spoof samples during fitting, \sysfull{} is close to that reference on ASVspoof 2019 LA ranking (8.45\% vs. 7.86\% EER), but it degrades more on 2021 DF ($+13.67$ vs. $+7.08$ EER points) and depends mainly on WavLM layer~3. A threshold taken from 2019 bona fide development does not remain valid under that shift: operational FPR rises to 20.23\%. Robust unlabeled recalibration restores the in-domain operating point up to about 10\% contamination. On 2021 DF the same rules restore the false-alarm budget, but TPR at 1\% FPR stays near 1.5\% even for an oracle target threshold. Later work should update the bona fide model under contamination constraints, rather than only retuning the decision threshold.
\section*{Acknowledgment}
R.S. expresses his gratitude for the funding support provided by the program ``Excellence initiative -
research university'' for the AGH University in Krakow, as well as the ARTIQ project (UMO-2021/01/2/ST6/00004 and ARTIQ/0004/2021), and by the funds of the Polish Ministry of Science and Higher Education assigned to the AGH University of Krakow.
\bibliographystyle{IEEEtran}
\clearpage
\bibliography{references_icdm2026_owad}

\end{document}